# High collection efficiency MCPs for photon counting detectors

D. A. Orlov,[*] T. Ruardij, S. Duarte Pinto, R. Glazenborg and E. Kernen

*PHOTONIS Netherlands BV,*
*Dwazziewegen 2, 9301 ZR Roden, The Netherlands*
*E-mail:* D.Orlov@nl.photonis.com

ABSTRACT: Multi Micro-Channel-Plate Photomultiplier tubes (MCP-PMT) with High Collection Efficiency (Hi-CE) MCPs are developed and characterised. With these Hi-CE MCPs more than 90% of photoelectrons emitted from the photocathode can be detected; this is in contrast to conventional MCPs where about 50% of photoelectrons are lost at the MCP stage. The drawback of the Hi-CE MCPs is a small degradation of the transfer time spread (TTS). However for applications where no sub-ns time resolution is required the implementation of Hi-CE MCPs is extremely beneficial, as it improves the detection efficiency of the MCP-PMT almost by a factor of two.

---

[*] Corresponding author.

# Contents

## 1. Introduction

A Micro-Channel-Plate photomultiplier tube (MCP-PMT) is a vacuum device consisting of a photocathode, absorbing the light and emitting photoelectrons, one, two or more MCP(s), multiplying the electrons from the photocathode by a factor of $10^5$-$10^6$ and an anode that can be single or segmented, to collect the output bunches of electrons. MCP-PMTs are widely used in scientific and medical applications for detection of low-intensity light signals, down to single photon counting requirements.

    The photocathode is the key part of the MCP-PMT which can be considered as the first stage of the amplifier. In particular, for Cherenkov radiation and LIDAR detectors, PHOTONIS has developed the S20 based Hi-QE photocathode series, improving the QE to above 30%, optimizing the QE-peak position within the spectral range 200-550 nm, and reaching dark count rates down to 30 Hz/cm$^2$ [1].

    The multiplication of a single photoelectron to a bunch of electrons is done by the MCPs, which have millions of channels working as a net of parallel Channeltrons. Electrons are accelerated along the channels and while striking the walls they generate secondary electrons. Repeated along the channel in a multi-stage multiplication, this process allows a very high gain ($10^5$-$10^6$) to be achieved within a sub-ns time window.

    The final value of the MCP gain is an essential parameter, however the collection efficiency (CE) of the first hit is demonstrated to play a major role, as it will allow a high value of Detective Quantum Efficiency (DQE). Indeed, for conventional MCPs with an open aspect ratio (OAR) of about 40-60%, the electrons hitting the MCP web are typically lost, resulting in a CE around 50% and thus a final DQE of half of the photocathode QE.

    In this paper we present results of CE measurements for conventional MCPs, and for MCPs with Hi-CE properties. We will demonstrate that the collection efficiency can be increased from 50% to almost 100% by using improved MCPs. The transfer time spread (TTS) is also

measured and demonstrates a small degradation of TTS for the Hi-CE MCPs; the expectation is that for experiments that do not require sub-ns properties, the drawback is negligible.

## 2. Experimental methods

The measurements are done on a specific type of MCP-PMT, PHOTONIS Planacon™ [2]. The chosen configuration is a 2 inch square MCP-PMT with a S20 Hi-QE photocathode, a dual 10μm pore MCP, and a 64 pixels anode. The reference Planacon™ has an MCP set with conventional geometry and electroding on the MCP entrance surface, the second Planacon™ uses a set of Hi-CE MCPs. In both cases the OAR of the MCPs is specified to be above 0.6 (optical measurement shows OAR to be 0.64).

For the gain and PHD measurements, the photocathode surface is homogeneously illuminated with a 400 nm LED source, of which the intensity is controlled by the LED current. An opal diffuser and neutral density filters are used to provide low-intensity homogeneous illumination of the photocathode. For calibration and stability control, the intensity of the scattered light is measured by a Si-photodiode. In order to calibrate the input photocurrents down to a few fA, first the photocathode is illuminated with a higher intensity, allowing to measure precisely a photocurrent in the order of 100 pA. Then the intensity of the light is reduced, and the reference Si-photodiode gives the proportional reduction factor on the photocurrent.

The MCP gain is measured in two ways, in DC- and in PHD-mode. For these measurements, the total signal from all 64 pixels of the anode is detected at once. The bias voltage from photocathode to MCP input is fixed to 200 V, the anode is set at the ground potential, and the MCP output is negatively biased to a few 100 V.

The DC gain is measured in the following way: the input photoelectron current, measured at the MCP entrance, is adjusted to about 100 fA and the DC photo-induced anode current (after subtraction of dark current) is recorded at different MCP voltages. The ratio of the resulting currents defines the MCP DC-gain. For all measurements, care is taken that the anode current stays below a few per cent of the strip current value in order to avoid saturation effects in MCPs.

The PHD-gain is calculated from the mean energy of the Pulse Height Distribution spectrum, where the measurements are done at a low level of illumination corresponding to a single photoelectron event. The PHD spectra is recorded using a Charge Sensitive Preamplifier CSP10 (1.4 V/pC), a shaping amplifier CSA4, and a multi-channel analyser MCA3 [3]. Calibration measurements show that the MCA scaling factor is 1.22 mV/ch.

The pulse height distribution of single photoelectron event (PHD) is measured in two steps. On one hand the pulse height distribution of the total signal (PHD-tot) is measured at a low illumination level where the photoelectron current arriving to MCPs is about 5 fA to 15 fA which corresponds to an input photoelectron rate of about 30-100 kHz. On the other hand, the background pulse height distribution (PHD-dark), which is typically just a few kHz, is measured with the LED light source being switched off, when the detected signal is only originating from the dark current of the photocathode with a small contribution of MCP dark noise. Finally, the light-induced pulse height distribution of single photoelectron event PHD is taken from the subtraction of background PHD-dark from PHD-tot.

To measure the collection efficiency of MCPs, the PHD data are used. The rate of the incoming flux of electrons to the MCP is adjusted as described above to provide the input rate in the range of 50-100 kHz while at the same time, the rate of the output pulses at the anode is measured using the PHD setup. The ratio of the input and output rates (after subtraction of the background) provides directly the value of the CE coefficient. The uncertainty on the CE-measurements is estimated to be about 3%; it is mainly originated from the accuracy of the photoelectron current measurements, uncertainty in counting of low-energy events in PHD curve, and from the statistical error of the detection rate.

To measure the Transfer Time Spread (TTS) of photoelectron bunches arriving to the anode, a 1 MHz laser pulses with a 100 ps pulse width is used [4]. To minimize the probability of generating a few photoelectrons per single laser shoot, the intensity of the laser is strongly reduced, achieving a generation rate of photoelectrons below 10% of the laser rate. The arrival time of electron bunches is recorded using a 2 GHz 20 dB preamplifier and a multiple-event time digitizer with a 100 ps time bin resolution (Model MCA6 [3]). The MCA6 has the capability to detect the rising and falling edges simultaneously, using the Constant-Fraction-Timing (CFT) mode. For TTS measurements only one anode pixel of the 64 present is used in order to minimize the noise in the detection line. The instrumental response function is about 230 ps (sigma) and is limited by the laser pulse width and the MCA6 resolution.

## 3. Results and discussions

In this section the measurements of DC- and PHD-gain are presented (section 3.1), as well as the collection efficiency data for conventional MCP and Hi-CE MCPs (section 3.2). In section 3.3 the timing characteristics of conventional and Hi-CE MCPs are studied and in section 3.4 the results obtained are discussed.

### 3.1 DC and PHD gains

Figure 1 shows DC and PHD gain measurements for Planacon™ devices with reference MCPs (top) and with Hi-CE MCPs (bottom) where both sets of MCPs have an OAR of about 0.64. The range of MCP voltages used is chosen to reach at least a gain of $10^6$. The blue and red curves are gain measurements done in DC mode and PHD mode, respectively (see section 2).

In the case of conventional reference MCPs, the DC-gain curve is below the PHD-gain on all the range of measured MCP voltages and has an average ratio of DC-gain/PHD-gain of about 0.58. This ratio is noticeably close to the optically measured OAR of the MCP (0.64).

The higher value of the PHD-gain compared to the DC-gain can be explained in the following way: in the PHD mode, only incoming electrons generating a bunch of electrons above a certain energy threshold are contributing in the distribution. Indeed the incoming electrons creating zero events or very low energy events are not detected, as they are below the small threshold used to filter out the measurement noise, and are thus not taken into account in the mean energy (gain) of all incoming electrons. In other words, only efficient incoming electrons are used to calculate the PHD gain. The difference between the PHD and the DC gain shows that approximately 40% of the incoming electrons is being treated as a lost event and is not contributing to the detectable signal.

For Planacon™ with Hi-CE MCPs (Fig.1 bottom) the PHD-gain values, measured with the same settings as in the reference case, are about the same as the DC-gain values, with an average ratio of 0.93. It proves that the Hi-CE MCPs are much more efficient, in comparison to conventional MCPs, to count almost all photoelectrons arriving to the MCPs. However, the accuracy of CE-measurements is estimated to be about 10% due to different scaling factors used to calculate mean energy (gain) from experimental PHDs. Further direct measurements of collection efficiency based on photon counting (see section 3.2) provide more accurate data.

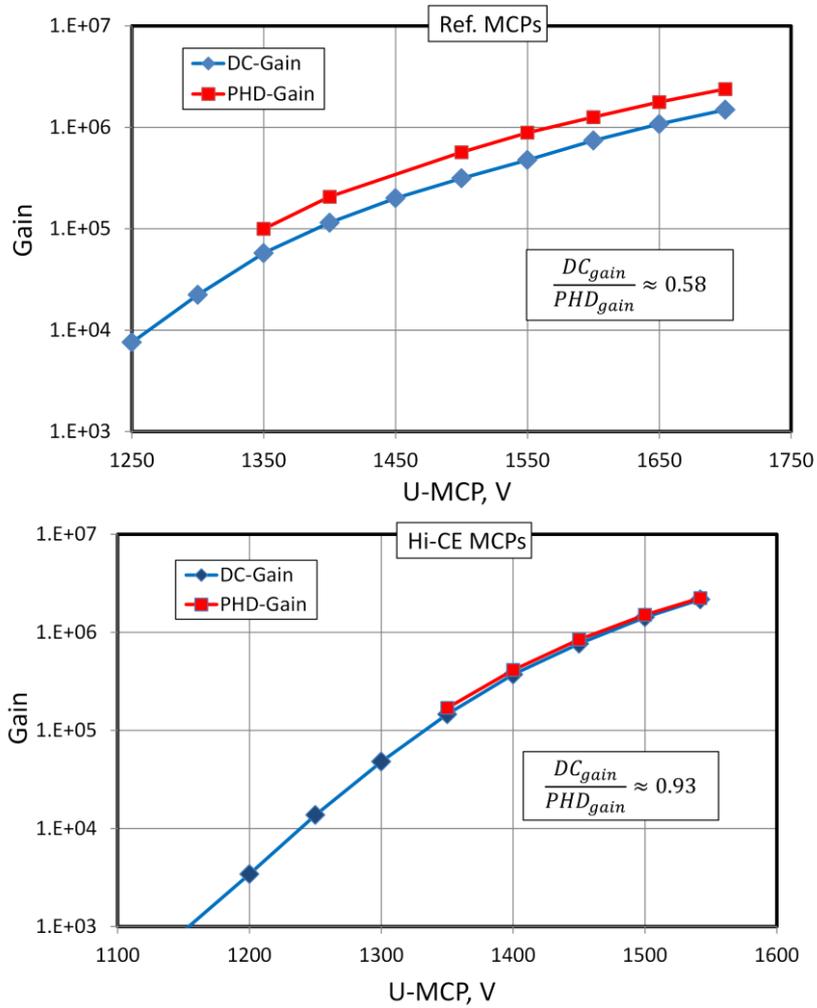

**Figure 1.** DC-gain (blue) and PHD-gain (red) vs MCP voltage for reference MCP-PMT (top) and for Hi-CE MCP PMTs (bottom).

### 3.2 MCP Collection efficiency

Figure 2 shows light-induced PHDs for Planacon™ with reference MCPs (upper figure) and Hi-CE MCPs (bottom). The contribution of the background PHD-dark is already excluded from the data and the input photoelectron rate is calculated as described in section 2. The values of input photoelectron rates and output rates are also shown on the figure.

     The total output rate is the sum of the counts for all channels of the PHD curve. One of the critical questions is how to calculate the counts of electrons close to zero where the contribution of the noise signal is rising. It has to be mentioned that the low-energy valley (defined partly by the measurement noise) is much below the PHD maximum, so the contribution of the PHD low-energy tail is small. Nevertheless to correct the PHD shape at low energies we set the channel threshold at about 25% from the peak position (that is far above the noise level) and we approximate the low-energy part, below the threshold, to be a linear extrapolation of the low-energy slope of the PHD curve.

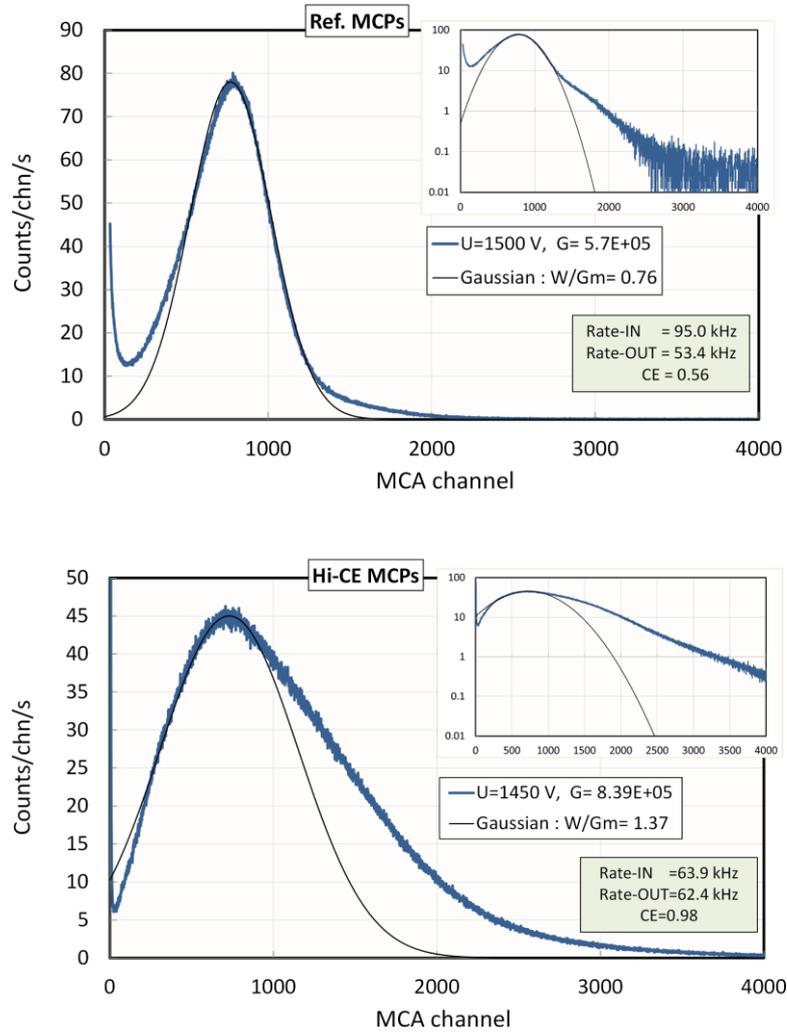

**Figure 2.** PHD-measurements used for rate counting of output pulses, with controlled input current/rate of incoming photoelectrons. The upper figure corresponds to MCP-PMT with reference MCPs, and bottom curve to Hi-CE MCPs. The black lines are the Gaussian fits of the top part of the PHDs. Inserts show the PHDs in log-scale.

The photoelectron input and output rates are shown in the insert in Fig.2 for reference and Hi-CE Planacons™. The collection efficiency of MCPs, defined as a ratio of output to input rates, is found to be 0.56 for ref-MCPs and 0.98 for Hi-CE MCPs. The measurements were also done at different MCP voltages and different input rates showing consistent values of CE-coefficient.

### 3.3 Timing performance

Figure 3 shows the temporal characterization of reference (top) and Hi-CE (bottom) Planacons™ for different photocathode-MCP extraction voltages. The main peak has a Gaussian shape with a $\sigma \cong 230$ ps mostly defined by the width of the laser pulse (100 ps) and the resolution of the multiple-event time digitizer with a 100 ps time bin. The TTS of the MCP-PMT in the Planacon™ configuration itself is expected to be in sub-100 ps range. Nevertheless, despite of the measurement set up limitation, the presented data give a clear view of the impact of Hi-CE MCPs configuration on the temporal characteristic of our MCP-PMTs, showing that these devices will present negligible degradation when no sub-ns resolution is required.

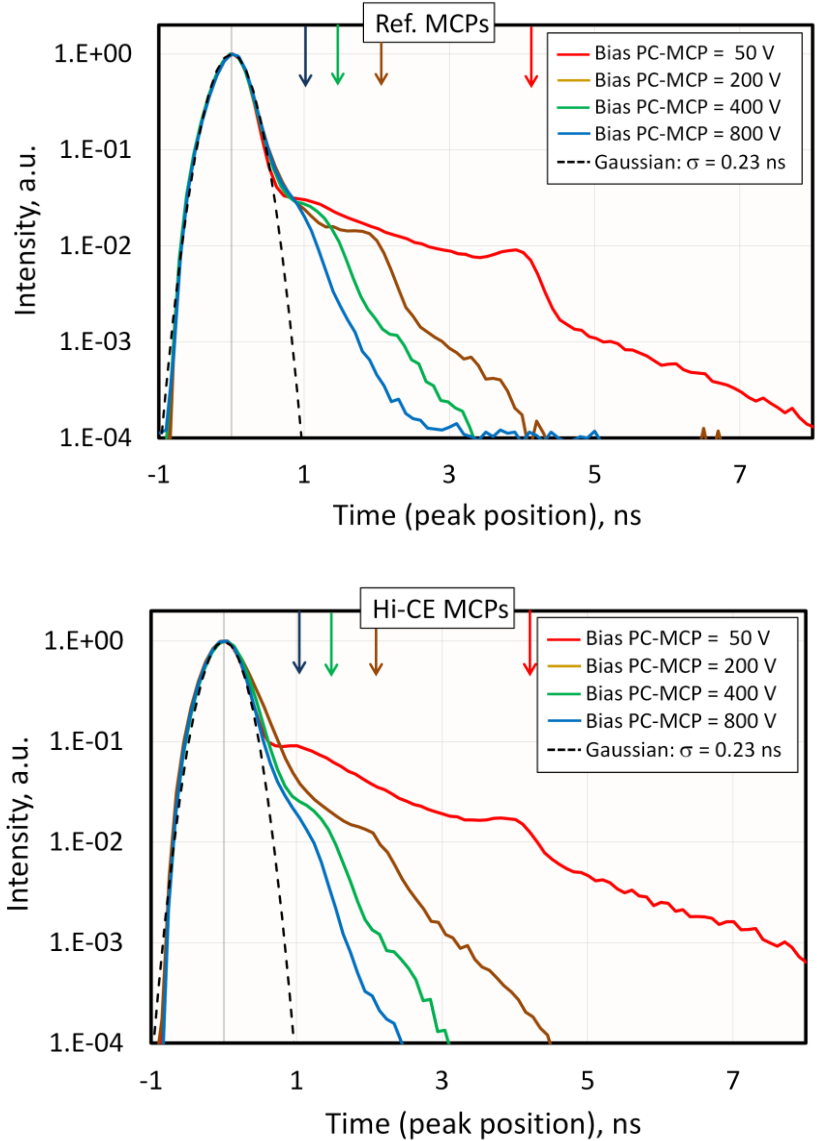

**Figure 3.** Temporal characteristics of reference (top) and Hi-CE (bottom) MCPs at different extraction photocathode-MCP voltages. The arrow shows the theoretical maximum delay of electrons elastically scattered from the front surface of MCPs.

One can see (Fig. 3) that in addition to the main Gaussian-like peak, the curves contain a high-energy tail which is due to elastically scattered, non-elastically scattered and secondary emitted electrons from the front surface of the first MCP. The arrows on the figures show the threshold for time position of elastically scattered electrons, calculated for different extraction photocathode voltages and taking into consideration the distance between the photocathode and the MCP (~4.3 mm). For each measured curve, the theoretical time position of scattered electrons corresponds to an additional feature along the high energy tail part of the spectra. The most clear example is in the case of low extraction voltage (50 V), where a peak can be observed at about 4 ns. The signal above these peaks can be assigned to electrons scattered a second time from the MCP surface. One can also see that the amplitude of the high-energy tail is about a factor 2 higher in the case of Hi-CE MCPs. Higher bias voltages push the right shoulder towards the main peak resulting in a decrease of the difference between reference and Hi-CE MCP.

Table 1 shows the contribution on the total output rate of electrons arriving after 0.5 ns and 1 ns, derived from the measured spectra, for reference and Hi-CE MCP-PMTs. At low bias voltage, this contribution is the highest and can reach 8% for a time window larger than 1 ns and up to 10 % for 0.5 ns for the reference MCP-PMT. When we switch to the Hi-CE MCPs the contribution increases by a factor 2. This drawback can be countered by increasing the bias voltage, as shown in the table where a bias voltage of 800 V reduces to about 1% the contribution of events being detected after 1 ns in both cases.

| Bias, V | Intensity above threshold, % | | | |
|---|---|---|---|---|
| | Ref. MCP | | Hi-CE MCP | |
| | >0.5 ns | > 1 ns | >0.5 ns | > 1 ns |
| 50 | 10.0 | 8.1 | 22.1 | 16.6 |
| 200 | 6.2 | 3.5 | 11.0 | 3.9 |
| 400 | 4.8 | 2.5 | 6.1 | 2.1 |
| 800 | 3.8 | 1.1 | 3.9 | 1.1 |

Table 1. The contribution of delayed electrons arriving to the anode after 0.5 ns and 1 ns of the main peak for reference and Hi-CE MCPs.

**3.4 Discussion**

The results presented above directly prove that the collection efficiency of ref-MCPs with conventional electroding is close to the OAR of the MCPs and in presented case is only about 56%. Almost all photoelectrons entering the web of the MCP are lost. In the case of Hi-CE MCPs, the situation is drastically improved; indeed almost all photoelectrons arriving at the MCP stage are generating pulses on the anode. The results presented were obtained with an optimized Hi-CE preparation process and have been tested on several tubes of different configurations. It is interesting to mention that even on MCP-PMTs using much lower OAR MCPs, a very high CE has been measured as well; CE close to 100% was obtained with an OAR of 0.49.

The benefit of the Hi-CE MCPs, compared to conventional MCPs, is almost a factor 2 increase of the Detective Quantum Efficiency (DQE) of the detector. However there are a few drawbacks that have to be taken into consideration. Due to counting of electrons scattered from MCP top surface (which are mostly lost for conventional MCPs) the shape of the PHD is slightly degraded, causing a broadening of the main peak and increasing the intensity of the high energy tail (see log-scale inserts in Fig.2). The FWHM to gain-at-peak ratio $W/G_m$ (see Fig.2) is found to be 0.8 for reference MCP and about 2 for Hi-CE MCP. This change of the $W/G_m$ is not a fixed factor, indeed in the presented case it turns out to be larger than typically observed with other tubes: there is an individual difference in PHD shape for each tubes. Despite of the broadening of the PHD for Hi-CE MCPs, the shape of the PHD is still good, with clear observed peak and small valley at low energies that should make no problem to detect the coming pulses.

As demonstrated in section 3.3 the timing behaviour of the MCP-PMT is affected by the Hi-CE MCPs. The presented data at 50 V extraction voltages shows the contribution of e-pulses delayed by more than 1 ns to be around 8% for conventional and 17% for Hi-CE. In both cases the contribution is high and it can be brought back down to a level of 1% for reference, as well for Hi-CE MCPs, when increasing the bias voltage to 800 V.

Another way to improve the timing properties could be to reduce the distance between the photocathode and the MCP which proportionally will shift the high-energy tail to the main peak and will decrease the contribution of delayed electrons.

When no sub-ns resolution is required Hi-CE MCPs can increase the Detective QE by almost a factor 2 when the MCP collection efficiency is close to 100%. To get information about temporal characteristic of Hi-CE MCP-PMTs within sub-ns range future work is required, including measurements with higher time resolution.

## 4. Conclusions

Direct measurements of collection efficiency of MCPs are presented and temporal characterization of conventional and Hi-CE MCP-PMTs are performed. While the collection efficiency of conventional MCPs is found to be about 50-60% and close to the OAR value of the MCP, it increases to about 100% for Hi-CE MCPs: almost all photoelectrons arriving to the MCP are being detected at the anode. The timing characteristic of the Hi-CE MCPs is slightly degraded with respect of conventional MCPs but when no sub-ns resolution is required, the implementation of Hi-CE MCPs is quite beneficial as it will improve the detection efficiency of the MCP-PMT almost by a factor of two.

## 5. Acknowledgements

We would like to thank Jack S., Henk V., Jan H., Hans de V., Ray F. and Sylvain P. for useful discussions and preparing the samples.